# Non-isothermal flow of Al-, Co- and Cu-based alloys made in different spatial configurations or structural states: model and experimental study

A.D. Berezner [a*], V.A. Fedorov [a], N.S. Perov [b], J.C. Qiao [c**], V.E. Gromov [d], M.Yu. Zadorozhnyy [e], G.V. Grigoriev [a]

[a] Theoretical and Experimental Physics Department, Derzhavin Tambov State University, Tambov 392000, Russia

[b] Magnetism Department, Lomonosov Moscow State University, Moscow 119991, Russia

[c] School of Mechanics, Civil Engineering and Architecture, Northwestern Polytechnical University, Xi'an 710072, China

[d] Department of Natural Sciences, Siberian State Industrial University, 654007 Novokuznetsk, Russia

[e] Laboratory of Functional Polymeric Materials, National University of Science and Technology «MISIS», Moscow 119049, Russia

Corresponding authors: * a.berezner1009@gmail.com, ** qjczy@nwpu.edu.cn

**Abstract**

The universal generalising approach for non-isothermal behaviour of different alloys has been provided together with the novel deformation modelling. Strong correlation between the model approach and experimental results is shown that permits estimation of main applied parameters such as the linear thermal expansion coefficient and others. Necking contours and critical thickness at corrugation for ribbon and rod specimens are also calculated. Fractal analysis of corrugation folds (their main size) has been carried out for polycrystalline and amorphous ribbon specimens. Structural peculiarities at the plastic deformation stage are investigated with microscopy.

# 1. Introduction

Study of elastic and plastic deformation is an essential, useful, and well-known approach in different fields of fundamental and applied science [1-6]. Each specimen exhibits a unique response depending on its initial and boundary deformation conditions [7]. Particularly, temporally increasing axial load leads to linear deformation of amorphous alloys up to their fracture [8], but the same conditions contribute to a noticeable strain hardening for polycrystalline specimens [9] together with further accelerating necking and final fracture. In addition, more complex (non-linear) stretching behaviour, such as those observed in auxetics [10], can occur due to chemical composition or interatomic arrangements. Another example of the system sensitivity to external conditions is a personal response to thermal impact. Isothermal loading mode for metallic alloys (i.e. heat transfer to them at non-room temperature) can lead to additional local reorganisations in the structure [11], and uniformly increasing thermal impact contributes to the change in the experimental curve [12]. Cyclic force loading at constant or variable temperatures [13,14] as well as switching between force vector projections (stretching or bending) [15] can distinctly affect deformation dynamics.

Isochronal dilatation regime [16] is qualitatively identical to thermal mechanical analysis (TMA) [17] due to similar deformation dynamics at static external loading [13]. In the first case, the primary driver of deformation is acting gravity, but at TMA, an automatic force preset, controlled by the bottom grip of a testing machine, is used. In both experimental modes, except for mechanical loading, additional acceleration of the process is provided by temporally uniform heating. Mentioned similarity in both deformation regimes is supposed to be properly analysed from the quantitative model standpoint because a complex study can simplify predictions of the sample response at different conditions (such as chemical composition, interatomic bonding, etc.). Note that non-isothermal behaviour remains theoretically understudied, despite its intensive experimental investigation [18]. Herewith, isothermal deformation has been properly investigated mostly at the elasto-plastic stage [19], where the exponential viscoelastic material model is used [7]. Moreover, in traditional (isothermal) creep theory [20-22], deformation is mainly considered for polycrystalline alloys (with respect to grain boundaries, dislocation mechanisms [23], and so on), which can be conceptually not effective (or even erroneous) for description of the deformation in amorphous structures (like alloys) due to their disordered structure [24]. In addition to dislocation models, various alternative approaches can be used to describe metallic glasses, including free volume [25], shear bands (or STZ) [26], flow units [27], quasi-point defects theory [28] or thermodynamic [29] ones.

During a study of plastic deformation and fracture, fatigue, or other tests, the general description of necking phenomena [30] (also corrugation in thin specimens [31]) is an important issue. At variable temperatures [32], the mentioned model analysis is complicated by non-linear boundary conditions, which require the necessary searching for a new concerted approach.

In connection with the noted non-isothermal deformation features, the main goal of this study is in the novel universal description of TMA, dynamic-mechanical analysis (i.e. DMA) and isochronal testing for Al-, Co-, and Cu-based amorphous or polycrystalline alloys. It will be achieved by cooperative use of thermodynamic [29] and deformation models [33-35]. Here we will consider plane and cylinder alloy configuration (ribbon and rod specimens, respectively), and deformation fracture features (such as necking and corrugation) are proposed to be investigated. Finally, a critical thickness of the specimens at starting corrugation and necking relief will be estimated.

## 2. Materials and methods

In this work, amorphous Al-Y-Ni-Co ($Al_{85}Y_8Ni_5Co_2$), Cu-Pd-P ($Cu_{54}Pd_{28}P_{18}$) and Co-Fe-Si-Mn-B-Cr (a set of widespread manufactured VAC6150, VAC6030, VAC6070, and MG2714 Vacuumschmelze/Metglass analogues) ribbon metallic glasses (*abbrev.* MGs) were chosen as testing specimens. Except mentioned MGs, we also used industrial polycrystalline Al-Fe (wt.%:

Al-98%, Fe-2%) ribbons and copper rods. All considered alloys are widely applicable in engineering and household tasks that are attractive to study their properties. Structural correspondence between investigated Cu or Al-Fe crystalline specimens and widespread analogues [36-37] was estimated with Bruker D2 Phaser (CuKα radiation) diffractometer (see Fig.1).

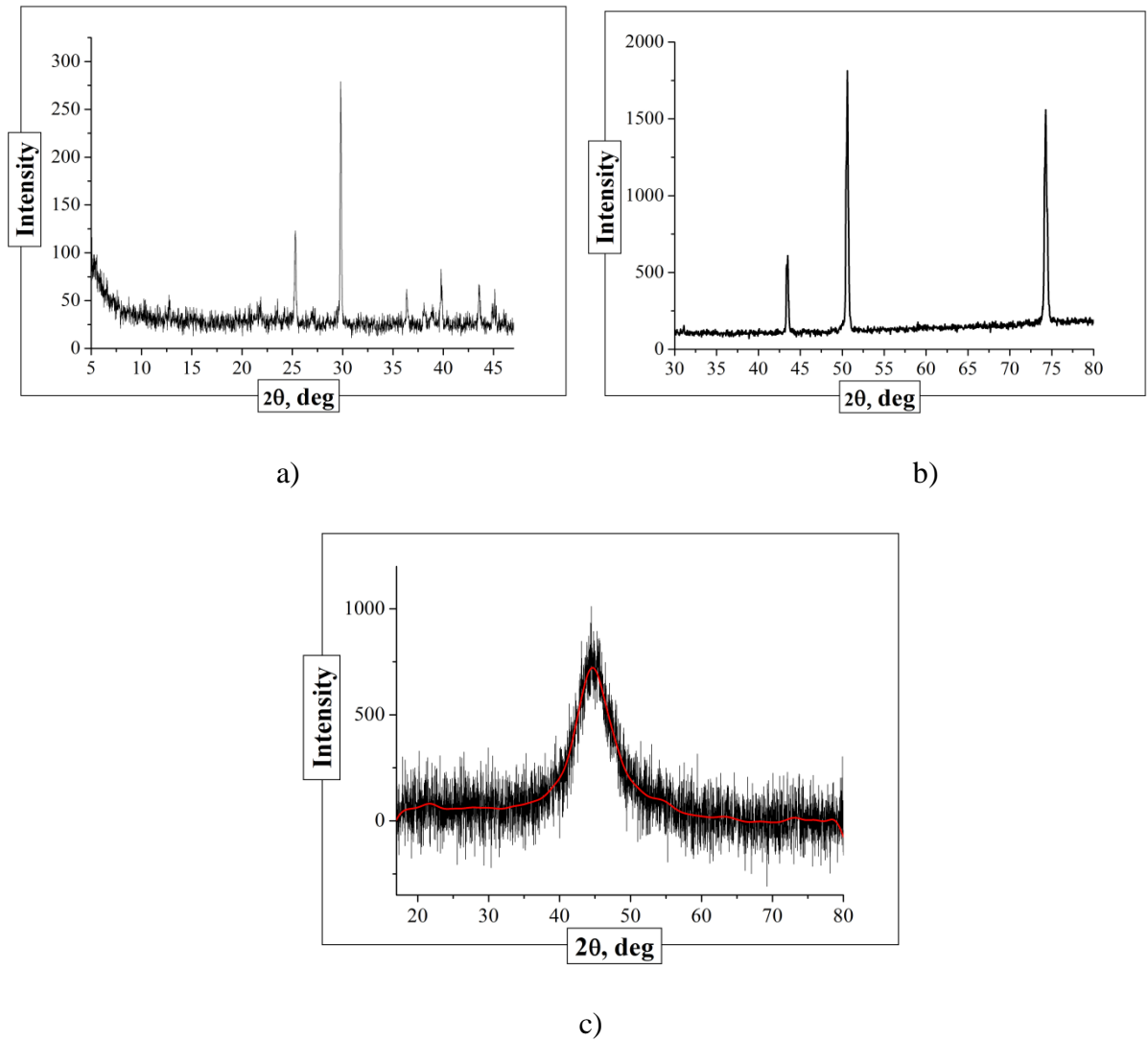

a)

b)

c)

Fig.1. X-Ray diffraction spectra for crystalline specimens: a) Al-Fe; b) Cu. Metallic glasses demonstrate smooth curves (i.e. without any peaks), and c) graph is one of them, for example. In the graphs, only useful intervals are plotted. Typical spectral lines of investigated crystalline specimens agree with standard crystal-lattice orientation data from literature

Thermomechanical analysis of Al-Y-Ni-Co and Cu-Pd-P ribbon metallic glasses was performed with a Q800 (TA Instruments) testing machine at 3 MPa external load and a 5 K/min heating rate up to 673 K. Working sizes of Al-Y-Ni-Co amorphous specimens were about 16.5 mm, 1.4 mm, and 0.02 mm (in length, width and thickness, respectively) with 16.5 mm, 2.5 mm, and 0.025 mm ones for copper-based MGs. DMA was also carried out with the Q800 (TA Instruments) device at similar experimental conditions (like TMA – by heating rate, scale, and composition) but in elastic harmonic oscillation mode (3 Hz loading frequency by 3 MPa plus static 12 MPa preloading) up to 540 K. Non-isothermal creep behaviour of polycrystalline Al-Fe or Cu alloys was studied using a Riftek triangle laser sensor in comparison with Co-Fe-Si-Mn-B-Cr MGs at 1 N of natural static gravity load (or 14 MPa stress) and 1 K/s of heating rate (a typical loading scheme is also provided in the **Appendix**). However, the sizes of Al-Fe crystalline ribbons were 50 mm in length, 6 mm in width, and 0.01 mm in thickness, while the copper specimens occupied 50 mm in length and 0.625 mm in diameter. For Co-Fe-Si-Mn-B-Cr MGs, nominal sizes were 50×3.5×0.02 mm.

Mentioned configuration (parallelepiped ribbons or cylindrical rods) was selected for the estimation of universal mechanical response in different specimens at the same axial testing conditions. We used the differential scanning calorimetry (DSC) with a 5 K/min heating rate for a detection the glass-transition ($T_g$) and crystallisation ($T_x$) temperatures of Al-Y-Ni-Co, Cu-Pd-P and Co-Fe-Si-Mn-B-Cr MGs as 533 K and 545 K, 528 K and 568 K, 716 K and 743 K respectively, that testifies to glass or crystal transition in some specimens during the testing. Also, the comparison between crystalline and amorphous alloys during the whole deformation experiment is required for better understanding the role of atomic reorganisations at non-isothermal impact.

For a surface relief estimation of finally tested specimens (after their two-fragment fracture), a FemtoScan (Advanced Technologies Center) scanning probe microscope (worked at the atomic-force, i.e. AFM mode) or Vega3 (Tescan) apart from JCM-7000 (Jeol) scanning electron microscopes (SEM) were used.

Magnetometry of Co-Fe-Si-Mn-B-Cr ferromagnetic MG was carried out with a LakeShore 7407 device at ±125–1000 Oe external magnetic fields (and ± 80 emu/g saturation magnetization) both in parallel and perpendicular projections relative to the specimens before and after deformation tests. In an additional experimental series, neodymium permanent magnets with the same field value (125–1000 Oe) were used for simultaneous induction of external field in non-isothermal creep tests.

## 3. Results and Discussion

### 3.1. Primary analysis of experiments and further generalization

In Fig.2., experimental TMA and isochronal deformation curves for ribbon and rod samples with different compositions and interatomic structures have been presented.

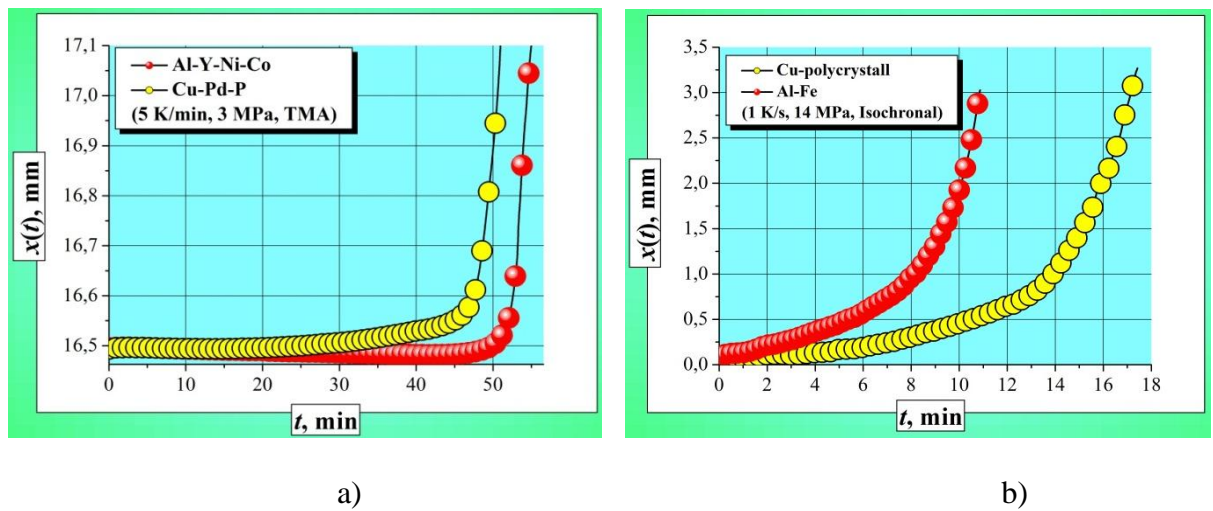

a) b)

Fig.2 Deformation curves of investigated specimens, plotted by: a) TMA data for Al-Y-Ni-Co and Cu-Pd-P amorphous ribbons; b) Isochronal testing for polycrystalline Al-Fe ribbons and Cu rods. DMA demonstrates the same deformation dynamics qualitatively as TMA

Analysis of the graphs in Fig.2 permits a conclusion about the identity (by $x(t)$ curve form) between the TMA, DMA, and creep. Herewith noticeable difference in the deformation rate (during TMA) can be caused both by structural features (MG or crystal) and by size-effect (copper rods are 25 times thicker than Cu-Pd-P ribbons). Despite the mentioned differences, interpolation by the empirical fractional function is possible for all experimental data in the form:

$$x(t) = x_0 + \frac{Ct}{B^2 - Bt}, \qquad (1)$$

with high correlation coefficient $r$ ($0.9 < r < 1$) and $x_o$ (or $l_o$), $C$, $B$, $t$ parameters, having physical meaning and stable dimension [29,33-35,38]. For TMA of Al-Y-Ni-Co amorphous alloys, relationship (1) has to be supplemented by the linear $-Dt$ term ($-D$ [m/s] is a process rate), which is related to reversible shrinkage of a specimen under the static load. All parameters, determined for the experiments, have been listed in Table 1, and model curves well coincide with ones in Fig.2 in the presented axis (micrometre) scale.

Table 1. Experimental parameters and their physical meaning in equation (1)

| Value<br>Composition,<br>Alloy type, Experiment | $C$ multiplier [$\times 10^{-3} \cdot 60$ m·s] | Fracture time $B$ [$\times 60$ s] | Shrinkage rate $D$ [$\times 10^{-3}$/60 m/s] | Linear correlation coefficient, $r$ |
|---|---|---|---|---|
| Al-Y-Ni-Co (MG, ribbon, TMA) | 0.43 | 55.2 | 0.00099 | 0.97 |
| Co-Fe-Si-Mn-B-Cr (MG, ribbon, isochronal test) | 3.9 | 15.5 | 0 | 0.997 |
| Cu-Pd-P (MG, ribbon, TMA) | 1.5 | 52.84 | 0 | 0.99 |
| Al-Fe (polycrystal, ribbon, isochronal test) | 6 | 12.6 | 0 | 0.998 |
| Cu (polycrystal, rod, isochronal test) | 8 | 19.5 | 0 | 0.996 |
| Al-Y-Ni-Co (MG, ribbon, DMA) | 1.3 | 52.61 | 0 | 0.98 |
| Cu-Pd-P (MG, ribbon, DMA) | 0.1 | 47.58 | 0 | 0.995 |

As one can see from Table 1, analytical parameters precisely describe deformation, and external experimental conditions impact them up to tenth and hundredth fractions of a number (more intensive loads and higher heating rates naturally reduce $B$ with increase of $C$). Moreover, Eq. (1) is a solution for an ordinary differential Duffing equation (see a proof in the **Appendix** here), which describes mechanical oscillations of a material point at non-linear external conditions [39]. These circumstances permit analysis of material behaviour via the same model despite different composition and geometry factors. Note that exactly controlling factors are heating rate and external loading because these parameters accelerate deformation by the way (1) with the appearance of a typical convex down $x(t)$ curve. Periodical loading (i.e. DMA) leads to the accelerated destruction of Al-Y-Ni-Co amorphous alloy that agrees with a calculation by formula (16) from [33] (which is also described as (A.1) in **Appendix**, here), and Al-based MG deforms three times slower at TMA compared with DMA. Herewith, the value order of the calculated deformation rate is about $10^{-4}$ s$^{-1}$. Likewise, for Al-based MG, the deformation rate of Cu-based MG, calculated for DMA, exceeds the same TMA parameter, and it can be explained by higher (at 1.6 times) loading at DMA. The presence of the $-Dt$ term (at TMA of Al-Y-Ni-Co) is caused by a reversible decrease in deformation rate, but it does not affect the main reaction force dynamics (as further differentiation of (1) with its using in Newton's law eliminates this term). Simultaneous induction of the external magnetic field does not noticeably change the deformation dynamics (i.e. non-isothermal creep by Eq. (1)) below or above $T_C$ Curie temperature (on average, $T_C = 600$ K for investigated Co-Fe-Si-Mn-B-Cr systems) despite the change in personal integral magnetic response, caused by collective spin-orbital relaxation [40] during the test (see Fig. 3).

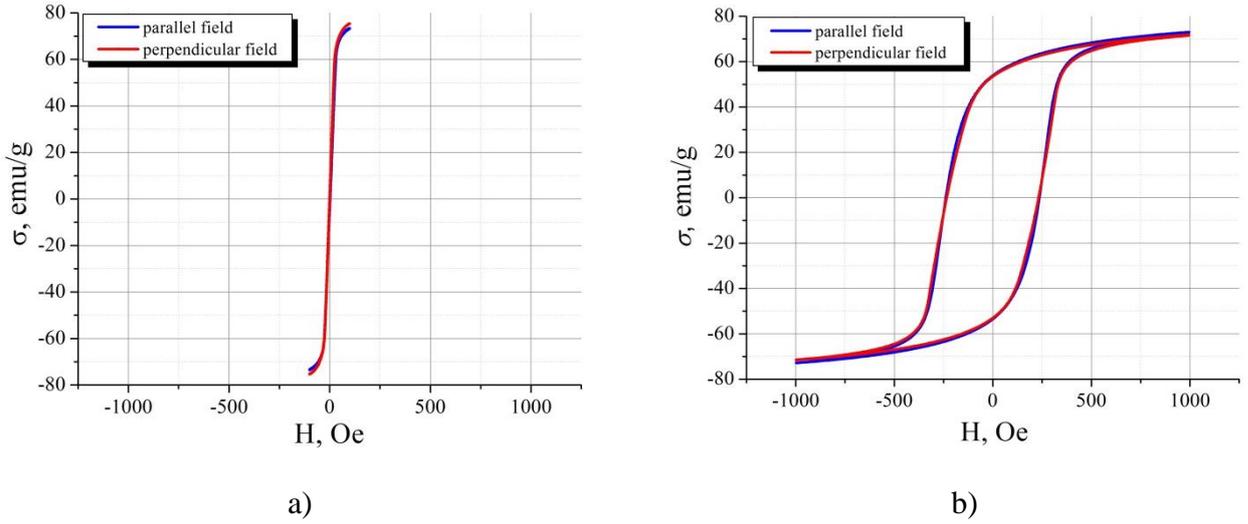

a)    b)

Fig.3. Typical hysteresis loops of a Co-Fe-Si-Mn-B-Cr alloy: a) at room temperatures, before a treatment; b) after non-isothermal creep testing up to 743 K with further cooling down to 300 K (the Curie temperature was 623 K)

We determined that alternative interpolation of all experimental graphs $x(t)$ by one or three standard exponent functions does not provide the necessary correlation (0.9 or more), as $r$ for exponent(s) is 1.5 times less than the same parameter from Table 1. Thus, equation (1) is universally applicable for deformation analysis of wide alloy range, and the necessary initial experimental conditions are: non-zero heating rate and positive (constant, oscillating or combined) mechanical loading. Deformation mechanisms of amorphous and crystalline materials are qualitatively similar due to the same stage when dislocations or shear bands form collectively identical flow structures. However, personal structural features such as beta-relaxation in MG [41] and others also can also appear during the experiment. Nevertheless, all differences in non-isothermal cases are quantitatively described just by the personal $C$ and $B$ material parameters in frames of the Duffing equation. Isochronal behaviour and TMA differ from isothermal deformation (with an exponential primary stage in the temporal strain curve) because of non-equilibrium thermo-mechanical reorganisation (in opposite to Maxwell or Kelvin-Voigt mechanisms, wherein the primary dynamics is caused by quasi-equilibrium structural changes). That is linearly rising heating with static external loading intensify only irreversible deformation of specimens stronger (without exponential primary stage) than the permanent heat transfer with fixed mechanical loading.

### 3.2. Force-function analysis

Calculation of the main physical and technical parameters for non-isothermal deformation using the values like in Table 1 is supposed to be possible. Due to this, some analytical relationships and calculated values are briefly expounded here (detail model development and basement have been described in the [29,33-34,38] articles). At TMA, the $A\sin(\omega t)$ harmonic term equals zero in the reaction force equation of a tested specimen (first derived in [38]):

$$F_{react.}(\omega; t(T)) = F_{load} - A\sin(\omega t) - \frac{2mC}{(B-t)^3} \qquad (2)$$

due to the fixed loading $F_{load} = mg$ [35] or just machine fixed $F_{load} = Const \neq 0$ one ($\omega$ cyclic frequency and $A$ amplitude are also equal zero, $m=10^{-4}$ kg is mass, $g$ is gravitational acceleration). During the action of linear temporal heating (with a constant $V_T$ rate), reaction force function (2) decreases, and a specimen finally breaks under the fixed load. With initial elasticity of specimens [38], using the relationship between material stiffness $k$ and its fracture time point $B$ ($k \approx 6m/B^2$ [34]), one can approximate $\delta$ phase lag behaviour analytically at $\varepsilon$ deformation, related to $\sigma$ stress during DMA (starting experimental interval only). In particular, using $\omega = \varphi/t = \sqrt{k/m}$ (where $\varphi$ is phase) classic wave equations with rough replacement (in them) of $\omega$ acting frequency by a similar material parameter (i.e. $\omega = \varphi/t \approx \delta/t$) provides $\delta = \omega t \approx t\sqrt{6}/B$ linear expression for $t$ time and $\delta$ phase lag in the elastic area. This estimation correlates with some part of the $\delta$-curve in Fig. 11 from [33] (between zero and $\beta$-relaxation times). Unlike the DMA, as additional periodical loading is absent, resonance and beating in the specimens [34] do not appear during TMA, that is well described by the model ($A\sin(\omega t)$ harmonic term is not added to the $\dfrac{-2mC}{(B-t)^3}$ fractional one). The temporal difference in atomic or molecular reorganisation ($T_g$ and $T_x$), occurring at DSC and DMA, decreases at TMA due to static load, but the impact of a various heating rate on structural relaxation remains. Absence of periodical loading also enhances structural viscosity that is described by less model terms (i.e. in [33], derivative of (18) and (19) by the $\dot{\varepsilon}$ rate has less negative terms if $\omega=0$; also Eqs. (18), (19) from [33] are mentioned in **Appendix** as (A.2) and (A.3), consequently). Moreover, in this case, viscosity remains non-Newtonian (due to linearly changing thermal activation). Generalising thermodynamic and statistical models [29] preserve their relevance and main relationships except for periodical terms. The model $C$ and $B$ values, depending on deformation dynamics, also can differ in each case.

### 3.3. Calculation of some material parameters for non-isothermal conditions

#### 3.3.1. Linear thermal expansion coefficient

Using the model from [42], the numerical value of the linear thermal expansion coefficient (CLTE) at TMA can be calculated for alloys via the derivative of (1) by temperature (i.e. $dx(t(T))/dT$ and $t = \dfrac{T - T_0}{V_T}$, where $T_0$ is the room temperature), which relates to time linearly. For Al-Y-Ni-Co MG (at TMA and $T$=507 K) $\alpha_L = 8.6 \cdot 10^{-6}$ 1/K, and crystalline Al-Fe one (deformation up to $T$=507 K) corresponds to $\alpha_L = 24 \cdot 10^{-6}$ 1/K that agrees by the magnitude order with the work [43] wherein Al-based ternary amorphous alloys were investigated at the same temperature. Similar numerical parameters, calculated for copper-based amorphous and crystalline alloys, also agree with literature data [44], and it varies between $4 \cdot 10^{-6}$ 1/K (MG) and $7 \cdot 10^{-6}$ 1/K (polycrystalline) in the temperature interval 300-500 K. For Co-based MG, $\alpha_L = 3.4 \cdot 10^{-6}$ 1/K (at $T$=300-400 K) and it also corresponds to Ref. [45] on multicomponent cobalt alloys. Note that CLTE tends to grow here for alloys with the same base component (for example, Al in Al-Y-Ni-Co and Al-Fe systems) because of different preset heating rates ($C$ and $B$ also depend on $V_T$), and alloy configuration can affect as $\alpha_L(Cu-Pd-P) < \alpha_L(Cu)$ with mainly thermal elongation of rods without lateral expansion or contraction inherent to ribbons.

### 3.3.2. Necking and fracture analysis for investigated alloys

Laminar and turbulent plastic flow of amorphous alloys were studied at axial experiments with different loads and heating [35,42,46]. Presence of viscous homogeneous deformation in ribbon specimens [33] permits using the reported model relationships (such as (1), (2) and others) for the further analysis of material necking and fracture with respect to possible nonlinearity. For estimation of necking form in the ribbon specimen, we use the relationships of elastic plane deformation with the linear heat transfer term [47]:

$$\begin{cases} \varepsilon_{xx} = \dfrac{\sigma_{xx} - \nu\sigma_{yy}}{E} + \alpha T \\ \varepsilon_{yy} = \dfrac{\sigma_{yy} - \nu\sigma_{xx}}{E} + \alpha T \end{cases}, \quad (3)$$

wherein $\varepsilon_{xx}$, $\varepsilon_{yy}$, $\sigma_{xx}$ and $\sigma_{yy}$ are diagonal components of strain and stress tensors respectively, $E$ is Young's modulus of material, $\alpha$ is CLTE, $\nu$ is Poisson's ratio, and $T$ is temperature. The expression of $\sigma_{yy}$ via other parameters (from the first equation in (3)) with its further substitution in the second equation permits to find a relationship between both deformation projections in the form (without the second repeating index):

$$\varepsilon_y(\varepsilon_x) = \dfrac{\sigma_x - (\varepsilon_x - \alpha T)E - \nu^2 \sigma_x}{\nu E} + \alpha T. \quad (4)$$

The further substitution of all material constants in Eq.(4) usually gives a linear deformation contour with ~ 45° angle that is well-known for traditional stretch tests. However, those values change during experiments (such as mentioned CLTE, temperature with $\varepsilon_x \sim x(t) \sim l(t)$), and consequently, we must use corresponding temporal and temperature functions to Eq.(4) with a

$$t = \frac{B^2 \Delta l}{C + B\Delta l} = \frac{T - T_0}{V_T}$$ substitution from [33] and additions $v \to 0$ if $t \to B$ instead the fixed numbers $\alpha, T, E, v$. It specifies the plastic flow contour (necking) at the final experimental stage (i.e. not only the elastic one). As precise Poisson's ratio dynamics is not known in our experiments (except initial number), this value is used for the second variable (with the temperature). Hence, in Fig.4, the main (schematic) view of possible necking profiles $\varepsilon_y = f(x;v)$ is plotted.

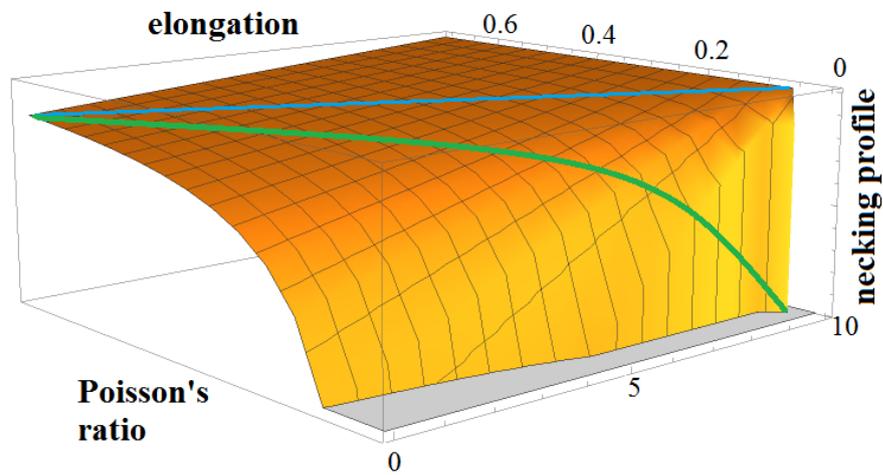

Fig.4. Surface of potential deformation profiles in a specimen. Elastic and viscous contours are mentioned by blue and green lines respectively

As one can see from Fig.4, among others, the non-linear necking profile (a projection on the «*f(x;y)*-elongation» plane) is possible that is approved by the SEM and optical microscopy (Fig.5).

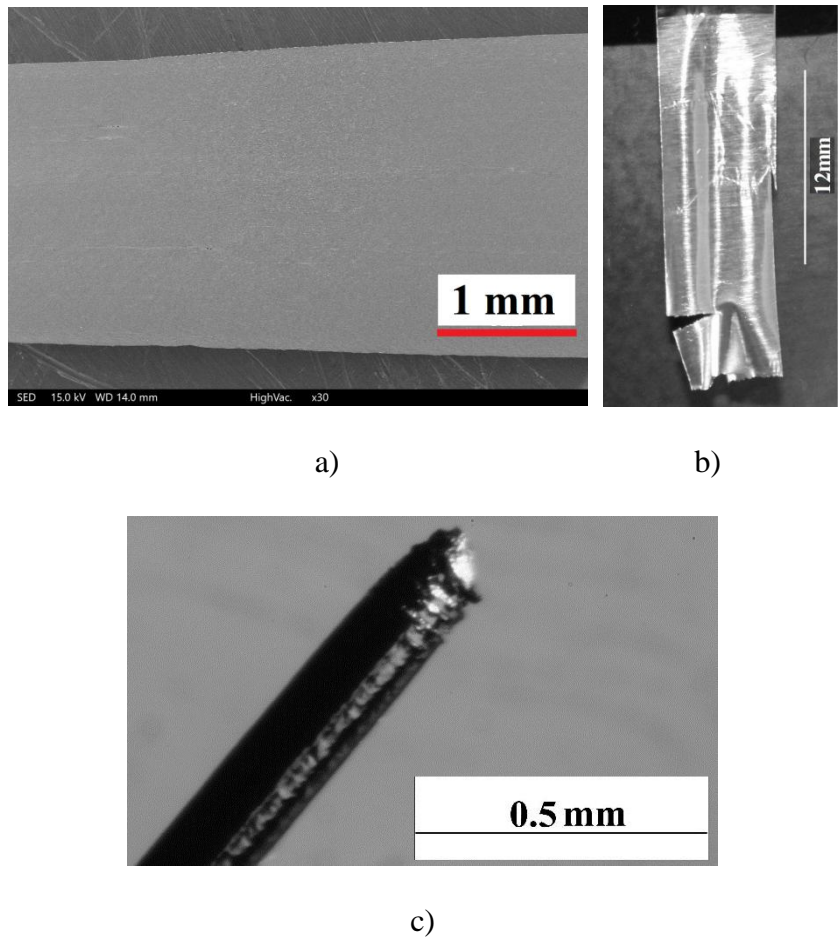

a) b)

c)

Fig.5. Fractography of specimens after TMA and static load: a) Cu-Pd-P ribbon MG (linear profile); b) Al-Fe ribbon (non-linear one); c) polycrystalline copper rod (non-linear «cigar-shaped» edging contour)

Necking contour of rods (cylinders) also can be considered in frames of Eq. (4) with the similar result because system (3) is rewritten in the cylindrical coordinates, if necessary. From Fig. 5b, the corrugation of a ribbon, appearing due to step-by-step crack opening from edge deeper to the centre, is also seen. The system (3) cannot be used for a description of this effect as its derivation is carried out in continuum mechanics. Therefore, it will be necessarily analysed below (in section **3.3.3 Corrugation and its analysis**). Note that for deformation of the type (3), power expression between the $\sigma_{yy}$ lateral stress and $\varepsilon_{xx}$ longitudinal strain (depending on transverse crack contour length) takes place. Moreover, the equation like fatigue Paris–Erdogan one [48] and others can be derived in the mentioned conditions. For this purpose, we use the first equation in (3) with the further substitution of deformation rate instead $\varepsilon_{xx}=\varepsilon_x$ (via recurrent equation $\varepsilon_x = \sqrt{\dfrac{C\dot{\varepsilon}_x}{x_0} - \dfrac{C}{Bx_0}}$ between $\dot{\varepsilon}_x$ and $\varepsilon_x$ from the work [35]). Finally, the required $\dot{\varepsilon}_x(\sigma_y)$ equation will be obtained as:

$$\dot{\varepsilon}_x = J - b\sigma_y^2 \sim \sigma_y^\beta, \qquad (5)$$

where $J$ and $b$ are constants, and $\beta$ is a power. As seen from (5), a fracture crack growth occurs at TMA and isochronal experiments similarly to fatigue tests from the model standpoint. This conclusion is experimentally approved by SEM (Fig.6).

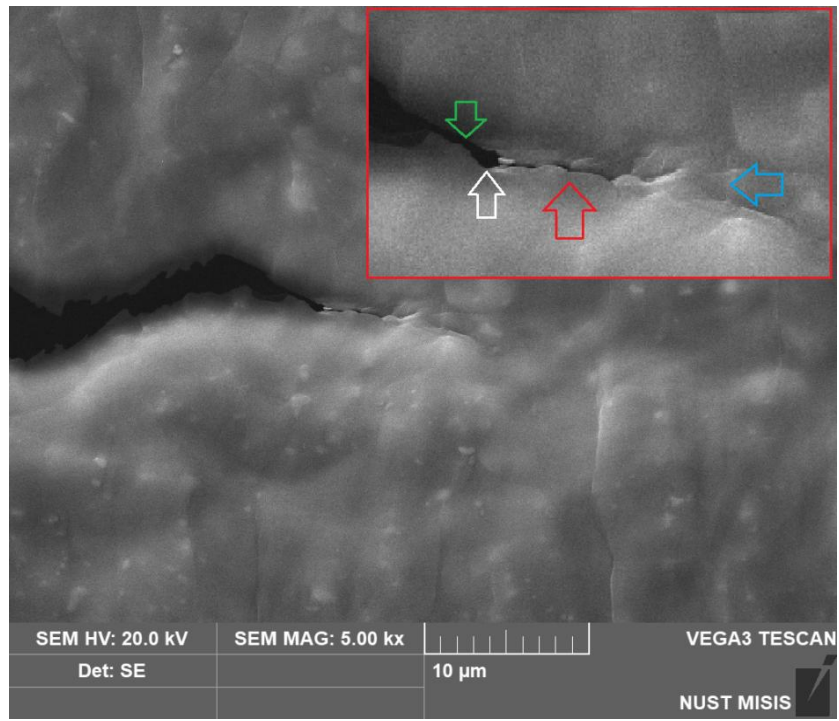

Fig.6. SEM image of a sector with the lateral crack in Al-Fe specimen after the unfinished test; an enlarged area near the crack tip is presented in the red insert with arrow pointers

In the crack propagation (see red insert of Fig.6), typical fatigue elements like a cavity (white arrow), ligament (green arrow), transgranular fracture with sharpening (red arrow), and angular stress striation (blue arrow) are observed [49].

### 3.3.3. Corrugation and its analysis

As we discussed in the previous section, system (3) does not describe the corrugation of a specimen because it contains a discontinuity (the growing fracture crack). That is why corrugation applied models like Euler–Bernoulli beam theory [50] and others cannot be used for the current task or they have to be combined as a step-by-step equation system for every moment of the fracture stage. Moreover, there are real overlapping (addable or subtractable) corrugated material waves, which are hard (also pointless or impossible) to describe by using the neutral axis approximation shifts constantly or breaks relative to $\varepsilon_y$ at the last deformation stage (up to the final fracture). Except for corrugation in MG (as in Fig. 7a), texturing (Fig. 7b), i.e. the forming of micro crack or pore netting with boundary (Fig. 7c) and complex relief (±1.3 μm level difference, see Fig. 7d), occurs in Al-Fe polycrystalline ribbons at TMA and static load deformation.

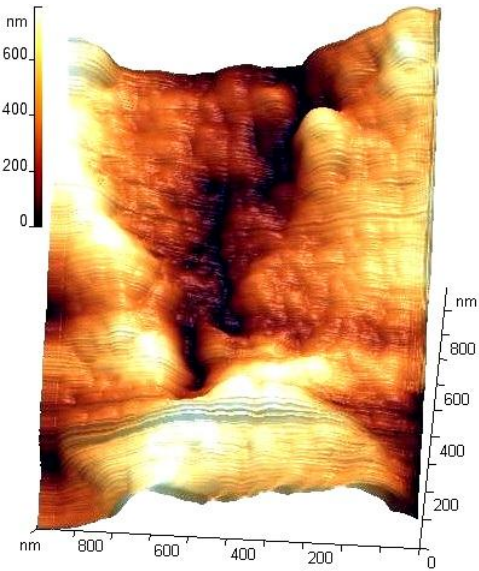 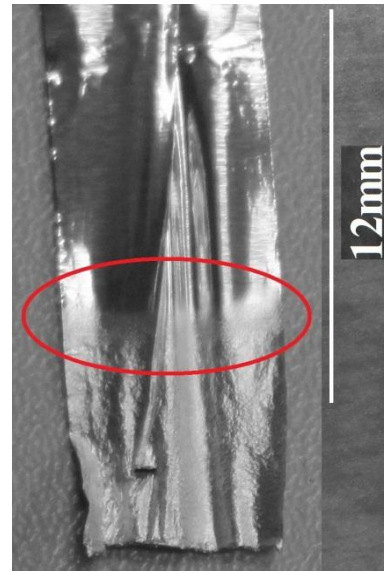

a)  b)

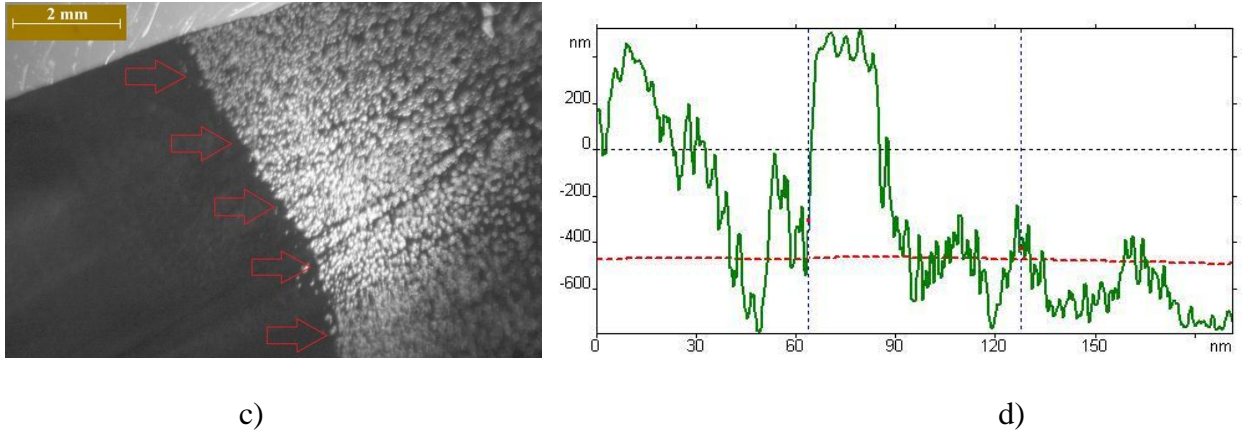

c)                                                d)

Fig.7. Corrugation and texturing of specimens at TMA and isochronal testing: a) An AFM surface of Cu-Pd-P MG; b) A part of the whole Al-Fe ribbon after the fracture (material preserves its new crinkled form), the transition boundary between smooth and corrugated areas is marked by the red oval; c) An optical microscopy photo, taken through transmitted light (the transition boundary was mentioned by red arrows); d) A selected AFM profile near the fracture line of an Al-Fe ribbon

At deformation of Al-Fe specimens, texturing can appear along coarse grain boundaries, or sub-fragmentation of the polycrystalline structure also is possible. As an alternative model approach, fractal analysis [51] and reference fractal dimension index $n$ can be used for a structural description of a ribbon specimen at its fracture stage (with corrugation). From this standpoint, the development of every fold is considered as percolation (progressive defect distribution), embracing some amount of particles in the specimen structure. By using the formula:

$$\xi \sim (P(T) - P_0)^{-n} \quad [51] \tag{6}$$

for calculation of a specific percolation element size (where $\xi$ is relative element size, $P$-$P_0$ is probability change), we estimate the corrugation scale. From [29], the probability distribution function (Eq. (14) from [29], mentioned here – in **Appendix** as (A.4), and others in [29]) at changing temperature and mechanical loading has to be used here in (6) instead $P(T) - P_0$, taking into account necessary experimental data (see Table 1). Then accounting for $n$~0.8 fractal index for glasses and amorphous polymers [52] gives the desired $\xi$ relative element size of $10^{-2}$ m at the 300-600 K temperature interval. Calculated $\xi$ value agrees by the magnitude order with the length of a ribbon and a longitudinal corrugation fold that testifies to self-similarity of forming folds and their nonlinear spreading.

As bulk specimens do not undergo corrugation (only brittle fracture occurs [53]) at the similar conditions [54], knowledge about the optimal (minimal) thickness without this effect is supposed to be necessary. Since the Reynolds number (Re) is the typical ratio for estimation of laminar and turbulent (nonlinear) flow or deformation regime [55], we use it here. Additionally, in the work [42], the temporal relation between this value and nonlinear deformation of ribbon amorphous specimens has been determined. And as the main expression, there was formula:

$$\mathrm{Re} = \frac{\dot{x}\rho s}{\eta} = \frac{m\dot{x}s}{V\eta} = \frac{m\dot{x}(t)}{ax(t)\eta(t)}, \tag{7}$$

wherein $a$ and $s$ were thickness and width, respectively, $\rho$ was density of a specimen, $\dot{x}$ or $\dot{x}(t)$ was deformation rate (as a derivative of (1) by time), $\eta$ was dynamic viscosity, $V$ was volume of a system. By substituting (1) and its derivative (instead $x(t)$ and $\dot{x}(t)$, respectively) into (7), the Reynolds number can be determined as a function of thickness and time, i.e. Re=$f(a;t)$. If we consider Re/Re$_o$ ~ 4000-6000 (under which specimen becomes to plastic flow [42]) at time $t$~$B$, the magnitude order of $a$ (at that Re/Re$_o$ acceleration is minimal) can be determined (see Fig.8).

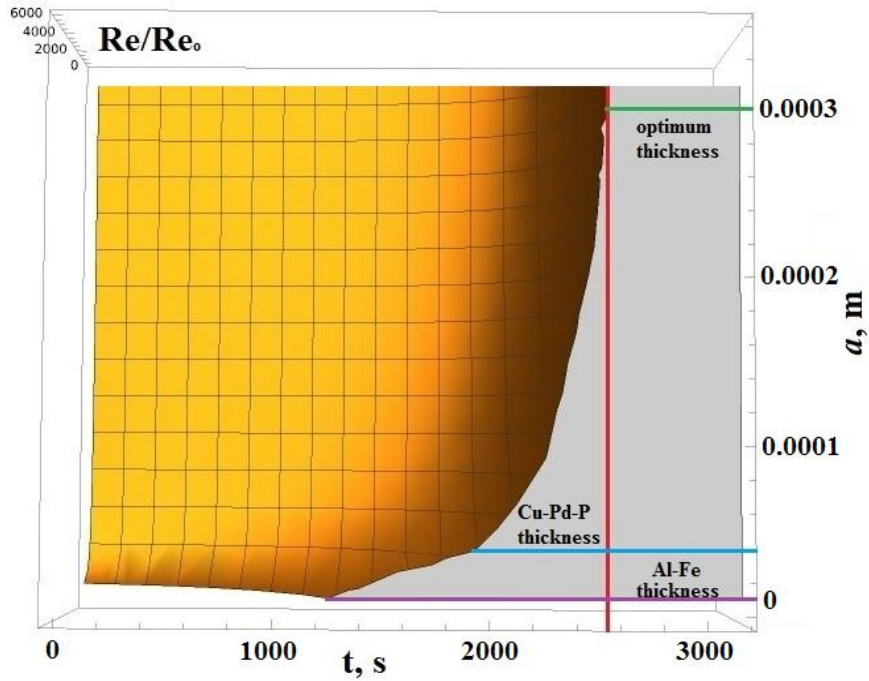

Fig.8. Re(*a*;*t*) surface. The boundary temporal point of the stable function area (along which, after 0.3 mm, the surface does not practically change) is mentioned by the red line. By green, blue, and violet lines, thicknesses are mentioned in the optimal case for Cu-Pd-P and Al-Fe investigated ribbons, respectively (Al-Y-Ni-Co and Co-Fe-Si-Mn-B-Cr specimens have the same thickness order as Al-Fe ones)

Thus, from (7), the optimal $a \geq 0.3$ mm thickness without corrugation can be found, i.e. the specimens must be 15-30 times thicker than considered ones if a more stable deformation is needed. It is known that ribbon (of 0.025 mm thickness) Cu-Pd-P specimens undergo corrugation yet [42] that agree with the calculated unstable flow area in Fig.8. Also, the diameter of using polycrystalline Cu rods is about 0.625 mm, at which we do not obtain any corrugation (fracture without visible longitudinal folds, i.e. different from the ones in Fig. 5b). In the graph (Fig.8), one can see the bottom (zero) limit of thickness wherein $Re/Re_o$ is not calculated due to its function gap that has the physical meaning and testifies to the adequacy (lower limit) of the proposed model.

## 4. Conclusions

Similarity between TMA, DMA, and isochronal testing exists, and it is described in frames of the same meso-physical model, using the non-linear Duffing equation for axial oscillations of a material point, despite the different specimen geometry or interatomic structure. Quantitative description is reached with model personal *C* and *B* parameters, providing the maximum correlation with the experiment. Calculation of important thermodynamic characteristics is possible together with necking and fracture analysis. All derived values and their numerical data are in good agreement with the experiment, and thus, the proposed model and approaches can be used for all other materials with the same deformation behaviour like (1) at mentioned conditions.

**Funding:** This research did not receive any specific grant from funding agencies in the public, commercial, or not-for-profit sectors.

**Declaration of competing interest**

The authors declare that they have no known competing financial interests or personal relationships that could have appeared to influence the work reported in this paper.

**Acknowledgements**

The results were partially obtained at the Center for Collective Use of Scientific Equipment at Derzhavin Tambov State University. Magnetometry was performed at Magnetic Materials Research Laboratory (Lomonosov Moscow State University).

**Appendix**

$$\dot{\varepsilon} l_0 B^2 \approx C, \qquad (A.1)$$

where $\dot{\varepsilon}$ is the deformation rate.

$$\tau(\dot{\varepsilon}) \approx \frac{F_{load} - A\omega B + A\omega\sqrt{\dfrac{C}{l_0}}\dot{\varepsilon}^{-1/2} - 2m\sqrt{\dfrac{l_0^3}{C}}\dot{\varepsilon}^{3/2}}{hl_0\sqrt{\dfrac{C}{l_0}}\dot{\varepsilon}^{1/2} - \dfrac{hC}{B} + hl_0}, \quad (A.2)$$

$$\tau(\dot{\varepsilon}) \approx M - a_1\dot{\varepsilon} - a_2\dot{\varepsilon}^{1/2} - a_3\dot{\varepsilon}^{-1} - \theta(\dot{\varepsilon}; a_i), \quad (A.3)$$

where $h$ is the thickness of a specimen. $a_1 - a_3$ with $M$ are division factors, and $\theta(\dot{\varepsilon}; a_i)$ is a remainder ($i>3$, $i$ – natural number) at division of numerator by the denominator in (A.2).

$$\alpha_0 + (\alpha_1 + \alpha_2)\cdot \exp(\tau) + \alpha_3 \exp(4\tau) \sim f(T), \quad (A.4)$$

with $\alpha_k$ exponential coefficients ($k=0–3$) and $\tau = \displaystyle\int \frac{dT}{V_T\left(B - \dfrac{T - T_0}{V_T}\right)}$.

**Creep loading scheme**

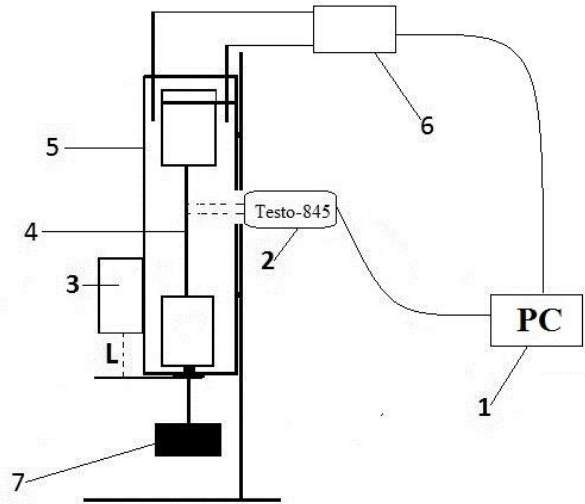

Loading scheme at creep testing: 1 PC, 2 the Testo-845 pyrometer, 3 laser sensor (measuring elongation), 4 specimen, 5 furnace, 6 temperature switcher, 7 static load

**Proof that Eq. (1) is a solution for the Duffing equation**

Function (1) is an exact solution of the $\ddot{x} - P_3(x) = 0$ (1a) or $\ddot{x} + c_1 x + c_2 x^2 + c_3 x^3 = 0$ (1b) non-linear homogeneous differential equation, where $P_3(x)$ is a cubic polynomial by $x$ variable with $c_1 = -6/B^2$, $c_2 = -6/CB$, and $c_3 = -2/C^2$ constant coefficients at $x$, $x^2$, and $x^3$ positive terms, respectively, and $-2C/B$ is the free term.

To prove this statement, the second temporal derivative of (1) with the further exchange from $t$ to $x$, using the inverse $t = \dfrac{B^2 \Delta x}{C + B\Delta x}$ [33] function relative to (1), should be calculated. Then, the derived expression for $\ddot{x}(t(x))$ is substituted into equation (1a) or (1b), and obtained fractions are algebraically simplified by reducing to a common denominator and performing mutual subtraction of opposite terms. In result, (1a) or (1b) becomes to a 0=0 identity.